\documentclass[conference]{IEEEtran}
\pdfoutput=1 
\usepackage{cite}
\usepackage{amsmath,amssymb,amsfonts}
\usepackage{algorithmic}
\usepackage{graphicx}
\usepackage{textcomp}
\usepackage{xcolor}
\def\BibTeX{{\rm B\kern-.05em{\sc i\kern-.025em b}\kern-.08em
    T\kern-.1667em\lower.7ex\hbox{E}\kern-.125emX}}
\begin{document}

\title{Generative AI and US Intellectual Property Law\\
}

\author{\IEEEauthorblockN{Cherie M Poland}
\IEEEauthorblockA{\textit{Complex Adaptive Systems Research and Virginia Tech} \\ 
Spring, Texas, USA \\
cmpoland1030@gmail.com \\
cheriem@vt.edu\\
ORCID:0000-0002-6345-649X}

}

\maketitle

\begin{abstract}
The rapidity with which generative AI has been adopted and advanced has raised legal and ethical questions related to the impact on artists rights, content production, data collection, privacy, accuracy of information, and intellectual property rights. Recent administrative and case law challenges have shown that generative AI software systems do not have independent intellectual property rights in the content that they generate. It remains to be seen whether human content creators can retain their intellectual property rights against generative AI software, its developers, operators, and owners for the misappropriation of the work of human creatives, given the metes and bounds of existing law. Early signs from various courts are mixed as to whether and to what degree the results generated by AI models meet the legal standards of infringement under existing law.
\end{abstract}

\begin{IEEEkeywords}
generative AI, law, copyright, patent, LLM, ethics, scraping, privacy, open-source, machine learning.
\end{IEEEkeywords}

\section{Very Slowly Then All-At-Once} 

Technically speaking, generative AI has been around since 1966 with MIT's ELIZA chatbot \cite{b1}, but it wasn't until the release of OpenAI's ChatGPT and Dall-E, Stability AI's Stable Diffusion, and Midjourney Inc.'s Midjourney in the fall of 2022 that the field of generative AI exploded onto the public square with wide adoption. This occurred even as Meta AI's Galactica chatbot was released only to be quietly withdrawn three days later because of its inability to discern truth \cite{b2}. In the rush to join the party, others introduced large language model (LLM) generative systems into the marketplace. Microsoft introduced Bing and Google introduced Bard in early 2023 \cite{b3}. Then, as if by surprise, the world realized how costly a generative AI misstep could be when Google shares dropped \$100 billion in one day (9\% in value for parent company Alphabet, Inc) after Bard produced a factual error in its first demonstration \cite{b4}.

Since then, Google has incorporated Generative AI into its search engine optimization with a Search Generative Experience \cite{b5}. The computer vision generative AI models have been joined by many others, while Midjourney has worked tirelessly in the background to improve its content results, such as eliminating the more-than-five-fingers seen in many image of generative hands \cite{b6}. Improvements will occur as a function of time, but at what cost to the to society and public policy?

\section{US Patent Law}
Both the US Patent and Trademark Office and the US Copyright Office have found that AI systems are not natural "persons" and as such cannot be inventors. Work generated by AI systems is not eligible for patent or copyright protection in the United States, even though patent and copyright protection may be available for humans who have created work within the same material containing work generated by an AI system. However, the extent of legal protection only reaches as far as the extent of the human contribution. These determinations are not surprising in light of the legal history and existing law.

The United States Court of Appeals for the Federal Circuit upheld a decision of the United States Patent and Trademark Office stating that the United States Patent Act requires an "inventor" to be a natural person \cite{b7}. The United States Supreme Court denied a Petition for Certiorari in the case fully rejecting computer scientist Stephen Thaler's attempts to seek patents on two inventions he said his DABUS system created \cite{b8}.

\subsection{Risks to the US Patent System}
Generative AI is already changing the computing and business landscape in ways that will supercharge the way machine learning (ML) and deep learning (DL) has been changing the entire ecosystem for the last decade since Ian Goodfellow's paper on Generative Adversarial Networks (2014) reignited the field of AI by reducing the complexity of the math-in-the-middle \cite{b9}. 

Early ML work using pre-LLM tokenized NLP models started a data-mining revolution in the assessment of emergent technology fields and in patents \cite{b10}. Similar work has been conducted on the use of predictive analytical tools to assess technological emergence and acquisition targets \cite{b11}.
This kind of research can be produced in a matter of hours by Generative AI systems from anywhere on the planet by mining public information and data stores to find trends.

This data-mining bridge to the future is also a two-way street. For all the good intentions and potential for new innovation, there is also a darker side that risks using public data against public institutions, for example by mining the public patent data bases and then generating gap-filling applications in areas between existing prior art.
This is not science fiction. It is a technique that has already been used in the US by large entities using teams of humans (often patent attorneys and agents) along with big data resources to identify gaps in existing patent portfolios and in the public patent portfolios of competitors. Such actions have historically had a very real institutional impact on the United States Patent and Trademark Office (USPTO) as it has had to redirect limited examination resources in order to address the influx of new applications. However, such behavior isn't illegal or unlawful at the present time.

Now with the speed of generative AI, imagine what could be done. If the use of generative AI is merely as a tool in the search-and-find process while the inventors of claims are human, then this may well circumvent the current USPTO policy and Court opinions, prohibiting invention by a generative AI system. However, to date, the agency has been silent on the use of generative AI as an assistive tool in the scientific discovery process.
The US Copyright office has required submissions of AI-assisted work to be disclaimed insofar as the generative AI portion is concerned, but it has not disallowed the submissions entirely.

In a world where a 100 page patent specification can be generated in a matter of minutes, reasonably complete, by a generative AI system, the only limiting factor to this potential emerging institutional risk becomes the fees for examination.

\section{US Copyright Law}

The US Copyright Office has determined that works created by AI without human intervention or involvement cannot be copyrighted because they fail to meet the human authorship requirement.
On March 15, 2023, the US Copyright Office issued a policy statement that works created with the assistance of artificial intelligence (AI) may be copyrightable, provided the work involves sufficient human authorship \cite{b12}. The Copyright Office has also required that copyright applicants disclose when their work includes AI-generated material, according to the Notice, and previously filed applications that do not disclose the use of AI must be corrected.

\subsection{Digital Millennium Copyright Act (DMCA)}

The Internet Digital Rights provisions, including Circumvention of Copyright Protection Systems, part of DMCA, \cite{b13} only applies to "persons" according to the plain terms of the statute. Thus, it is likely any DMCA challenges to infringement by generative AI systems will fail under the same basis and for the same reasons that the USPTO and the Copyright Office have found generative AI systems to be non-inventors and non-creators. Generative AI systems are not "persons".
However, under the common law of agency, as well as theories related to alter ego and piercing the veil, the individual humans, companies, and related entities behind the generative software can be held accountable for the acts performed by the software. Additionally, DMCA claims are not predicated on infringement claims, rendering the potential affirmative defense of fair use, inapplicable \cite{b14} (explaining fair use as an affirmative defense).

\subsection{Monkey Selfies}

In Naruto v Slater (9th Cir. 2018) \cite{b15} the court affirmed the district court's dismissal of claims brought by a monkey, Naruto, stating that although the animal had constitutional standing, it lacked statutory standing to claim copyright infringement of photographs known as "Monkey Selfies."

\subsection{Kashtanova and Zarya of the Dawn}

In February 21, 2023, the US Copyright office held that although the Zarya of the Dawn "work" was co-authored by Kristina Kashtanova and Midjourney's AI technology, the contributions of Ms. Kashtanova could be protected by copyright, including the text of the work, the selection, coordination, and arrangement of the work's written and visual elements, even thought the images in the work were generated by the Midjourney technology and are not the product of human authorship \cite{b16}. In order for Ms. Kashtanova's contributions to be protected, the Copyright Office also required that she disclaim the Midjourney-generated content.

\subsection{The Third Compendium}

In the Third Edition of the United States Copyright Office's Compendium of U.S. Copyright Office Practices, 28 January 2021, is dispositive on the issue of the requirement for human authorship \cite{b17}.

\subsubsection{Section 306} states
    "The U.S. Copyright Office will register an original work of authorship, provided that the
    work was created by a human being. The copyright law only protects “the fruits of intellectual labor” that “are founded in the creative powers of the mind” \cite{b18}. Because copyright law is limited to “original intellectual conceptions of the author,” the Office will refuse to register a claim if it determines that a human being did not create the work \cite {b19}.

\subsubsection{Section 313.2} states
    "The Office will not register works produced by a machine or mere mechanical process that operates randomly or automatically without any creative input or intervention from a human author. The crucial question is “whether the ‘work’ is basically one of human authorship, with the computer [or other device] merely being an  assisting instrument, or whether the traditional elements of authorship in the work (literary, artistic, or musical expression or elements of selection, arrangement, etc.) were actually conceived and executed not by man but by a machine.”

\subsection{Limitations of Fair Use and Transformative Use}

US copyright law retains limits on the exclusive right to reproduce copyrighted work as set forth in the Copyright Act of 1976, as amended. However, it also permits transient copies made during computational processing (cache) without the need to resort to fair use.

There are some who argue that AI researchers maintain the right to conduct computational research on literature and using publicly available images, even when the material is copied from an infringing source. The public policy argument in favor of the right to conduct computational research relates to social benefits in information sharing. However, this argument is not necessarily supported by the law.

In Andy Warhol Foundation for The Visual Arts, Inc. vs Goldsmith \cite{b20} the U.S. Supreme Court analyzed the first of the four fair-use factors, "the purpose and character of the use, including whether such use is of a commercial nature or is for nonprofit education purposes" and held the commercial nature of a use is relevant but not dispositive. "Fair use is an objective inquiry into what a user does with an original work" (Slip Op. 5). The Court reinforced the prior holdings in \cite{b14} as being instructive on the questions of fair use and transformative use. "Campbell describe a transformative use as one that “alter[s] the first [work] with new expression, meaning, or message” \cite{b14} at 579. "But Campbell cannot be read to mean that §107(1) weighs in favor of any use that adds new expression, meaning, or message. Otherwise, “transformative use” would swallow the copyright owner’s exclusive right to prepare derivative works, as many derivative works that “recast, transfor[m] or adap[t]” the original, §101, add new expression of some kind. The meaning of a secondary work, as reasonably can be perceived, should be considered to the extent necessary to determine whether the purpose of the use is distinct from the original" \cite{b20} (Slip Op. 6).
 
\subsection{Derivative Works}
Under US law, derivative works refer to copyrighted work that arises from another copyrighted work. Copyright allows owners to decide how their works can be used, including derivative works from the original \cite{b21}, \cite{b22}, \cite{b23}.
Digital version of artistic endeavors fall within the category of derivative works under US copyright law \cite{b14}.\\
In Campbell v. Acuff-Rose Music, Inc., \cite{b14} the US Supreme Court found that although a parody of the song "Oh, Pretty Woman" by 2 Live Crew was an unauthorized derivative work, fair use was available as a complete defense. The holding in Campbell \cite{b14}, was reinforced by the holding in Andy Warhol Foundation for The Visual Arts (18 May 2023) \cite{b20}. This seems to indicate that a fair use affirmative defense, even to derivative works may be available if the facts are in accordance with all of the articulated derivative, transformation, and fair use tests.

\section{Caveat Emptor: No Free Ride for Automation}

The act of web-scraping the Internet by automated means does not exempt one from potential copyright violations. Digital derivative works may be present online and the applicable law follows them, as the limitations on use still belongs to the copyright holder. hiQ Labs, Inc. v. LinkedIn Corp \cite{b25}. hiQ Labs v. LinkedIn Corp is the current prevailing US case on the limits of web scraping. In a November 2022 remand from the US Supreme Court, the Ninth Circuit ruled that hiQ had breached LinkedIn's User Agreement by scraping content from the site. Similarly, works with watermarks and licensing requirements may also be indicia that the work is copyrighted and should not be included in an AI dataset.

In October 2022, the US White House Office of Science and Technology Policy (OSTP) proposed a Blueprint for an AI Bill of Rights, a nonbinding suggestion for the responsible use of AI \cite{b26}. In the document, an automated system is defined any system, software, or process that uses computation as whole or part of a system to determine outcomes, make or aid decisions, inform policy implementation, collect data or observations, or otherwise interact with individuals and/or communities. Automated systems include, but are not limited to, systems derived from machine learning, statistics, or other data processing or artificial intelligence techniques, and exclude passive computing infrastructure." However, in a subpart, the policy also defines “passive computing infrastructure” as any intermediary technology that does not influence or determine the outcome of decision, make or aid in decisions, inform policy implementation, or collect data or observations, including web hosting, domain registration, networking, caching, data storage, or cybersecurity.\\
The codification of such a definition could someday boot-strap policy into legislation of generative AI systems to either give a pass to generative AI systems in their use of the creative works of others (similar to the copyright law exceptions in Japan \cite{b27}) or to classify generative AI systems as something other than the "passive computing infrastructure" given that some have controversially argued that the process of AI training of content entails merely a passive reference to content created by others. Is the process of training generative AI models an automated or a passive system under the OSTP definition? To what extent might that definition be used to apply existing intellectual property law to cases in controversy?

\section{Potential Harms and Mitigation} 
There are many possible harms along the generative AI continuum, from input problems related to data collection, scraping, consent, privacy, content, control of content, fairness, licensing, and copyright, to middle-model problems with model documentation, reporting, standards, accountability, and reproducibility, to output problems related to accountability, compliance, security, repositories, and legal liability.

The lack of uniform laws related to data, fairness, accountability, transparency, and intellectual property rights (IP) in the space around generative AI will continue to create tensions in the field, especially as legislators and regulators grapple with the "what-ifs" of how to response to actual and potential threats and public harms. The potential for public anger, disillusionment, concern about trustworthiness, abuse of the IP rights of others, and the lack of truthfulness (hallucinations) in the responses of generative AI systems is growing. The manner in which different jurisdictions handle public concerns and perception is something to watch as the wide-spread adoption of generative AI proceeds, along with the potential harms associated with hallucinations (false facts) and non-scalable reinforcement-learning with human feedback (RLHF).

In all cases, there is no free ride for automation. Software will not be held accountable by a court of law. Humans and legally-recognized business entities will. The limits of web-scraping were settled by the 9th Circuit Court of Appeals and the US Supreme Court in hiQ Labs, Inc. v. LinkedIn Corp \cite{b25}.

Courts will find humans, not software, responsible for bad conduct and behavior. Courts may or may not look to Terms of Service (TOS) and Terms of Use (TOU) as means by which companies can escape liabilities for their actions and for the known problems and known lack of resolution of the problems with their software models.

If regulators or courts decide to hold companies accountable under strict product liability or tort laws, the TOS and TOU may not matter. Entities providing generative AI models would do well to consider the historical nexus between strict products liability law and public harm, especially in countries like the United States. There is a lengthy history of accountability with automobiles and other advanced technologies within products liability law. Courts will not look at the technology in isolation. Rather they are more likely to consider the application of existing law as the more immediate and effective solution.

In the case of generative AI and authors and artists rights should look to the music industry and the licensing and rights management schema of BMI, ASCAP, and SESAC. 
In the case of TOS and TOU, contract, tort, and products liability law can all apply. In the case of physical or mental harm, tort law and even criminal statutes may be applicable.
 
Entities and individuals who permit technologically immature generative AI models with the potential for human harm, including models with high hallucination, disinformation, and deepfake potential, who fail to withdraw these systems from the public square when the generative AI systems engage in actions that humans would not otherwise be permitted to engage in, risk serious legal consequences. Similarly, generative AI should not, but will likely be used as a veil or as cover for bad actors, cybercrime, fraud, and malfeasance by humans. Generative AI is a technological tool. Those wielding it will bear the legal responsibility of making and using it.

\subsection{New Cybersecurity Risks}
At the RSA Conference held in April 2023, National Security Agency (NSA) cybersecurity director Rob Joyce was quoted as saying "[T]he NSA expects generative AI to fuel already effective scams like phishing." Additionally, "That Russian-native hacker who doesn't speak English well is no longer going to crafty a crappy email to your employees. Its going to be native-language English, its going to make sense, its going to pass the sniff test ... So that right there is here today, and we are seeing adversaries, both nation-state and criminals start to experiment with ChatGPT-type generate to give them English-language opportunities" \cite{b28}. Joyce also warned that "although AI chatbots may not be able to develop perfectly weaponized novel malware from scratch, attackers can use the coding skills the platforms do have to make smaller changes that could have a big effect. The idea would be to modify existing malware with generative AI to change its characteristics and behaviour enough that scanning tools like antivirus software may not recognize and flag the new iteration" \cite{b28}. "Its going to help rewrite code and make it in ways that will change the signature and the attributes of it," Joyce said. "That [is] going to be challenging for us in the near term" \cite{b28}.

\subsection{Attempts at getting around the law (aka circumvention)}
Some entities building and operating both computer vision (CV) and large language models (LLMs) appear to be engaging in various forms of legal circumvention due to limitations in existing licensing agreements.

The non-profit LAION group, who advocates for democratizing AI research globally, has been releasing open datasets, code and machine learning models \cite{b29}. They make models, datasets and code reusable without the need to train from scratch all the time. They also provide datasets of different size and focus. On its website, LAION states that its datasets are simply indexes to the internet \cite{b30}. That is, they are lists of URLs to the original images together with the ALT texts found linked to those images. In order to protect themselves from copyright violations, the LAION group has downloaded and calculated CLIP embeddings of the pictures to compute similarity scores between pictures and texts. They then subsequently discarded all the photos as a means of circumventing allegations of direct copying. However, any researcher using the datasets must reconstruct the image data by downloading the subset they are interested in. LAION suggests using the img2dataset tool for this purpose \cite{b30}.

Meta AI has apparently engaged in their own form of circumvention due to the terms of the original licensing agreements attached to their LLaMA models. LLaMA models that were supposed to be used for research purposes only, were leaked online shortly after their initial restricted release \cite{b31}. Meta AI has stated that they provide XOR weights for their open access models because their of strong copyleft license GPLv3 that extends to both the original LLaMA models and derivative versions does not permit them to distribute LLaMA-based models \cite{b32}, \cite{b33}. The XOR weights are the difference between the original LLaMA weights that have been trained by Meta AI and the fine-tuned weights trained by others \cite{b32}, \cite{b34}.

Subsequent users can further tune the LLaMA model for different tasks, like user interactions and use the resulting models as they please because one has to have the original XOR weighs in order to use them \cite{b35}. However, in order to get the original XOR weights, all you have to do is run a few lines of python inference code that is readily available on GitHub as an xor\_codec.py \cite{b32}, \cite{b34}.

There are legitimate concerns as to whether this legal-strawman will be a sufficient copyright law work-around. No court has specifically ruled on this issue. The schema relies on the exploitation of a gap in standard contract/licensing law by those developing and profiting from generative AI technology.

Using different type of circumvention, LAION has started including filters in its datasets for assessing and excluding (at the user's discretion) watermarked images and not-safe-for-work (NSFW) images and text \cite{b36}, \cite{b37}. As an example, the LAION5b dataset utilizes classifiers and nearest neighbor indices with detection scores and tags for image watermarks and NSFW content \cite{b38}. The curators of the datasets still need several disclaimers about the datasets, but the circumvention filters provide a starting-point for justifying a whole-of-Internet web-scrape by providing data-filtering as a feature of the dataset by optional selection/deselection of data. However, recent audits of the system have found that excluding the files marked NSFW are not fool-proof as there are many other scraped images in the dataset that may be considered NSFW by some, but perhaps did not meet the initial LAION5b cutoff for distinctive tagging.

\section{Conclusion}

The laws related to copyright, fair use, transformation, and derivative works remain applicable to material scraped from online sources that is then used by and in generative AI systems. The courts are going to have to decide whether generative AI technology falls into established intellectual property law and if so, to what degree. State and federal legislatures will then have to gap-fill with new legislation, in the interests of public policy, when the existing law falls short of its intended goal of protection of intellectual property rights. The process may be long and will likely be determined on a case-by-case basis at the outset, at least until one or more cases makes it to the US Supreme Court. However, courts and legislatures are not new to the challenges of technology. In the 1990s the challenges arose from questions about the metes and bounds of Internet jurisdiction, P2P distributed application architecture, and copyright as it pertained to file-sharing. Compliance with the law, licensing, and ethics is always a good policy.

\section{Future Considerations}
As far as the question of artificial general intelligence (AGI) is concerned, none of the current AI or generative models in existence are capable of AGI. All rules-based systems have their trade-offs under-the-hood and behind the proverbial curtain. These trade-offs in weights, biases, fit, optimization, and reward will continue to be be accounted for by engineers and developers as they have always been.\\
Insofar as autonomy (autonomous systems) is imputed to generative AI systems and AI in general, humans remain conflicted about their own bodies and control over them and by whom and under what circumstances. The question of who holds the ring-of-power for decisions of self-autonomy over humans is still very much a raw question. It has many different observations and outcomes in medical, medicament, and life/death contexts and varies greatly across cultures, countries, and religions.\\
Self-determination is a fundamental principle in international relations as it arises from a basic human right. Yet, laws and policies are often in conflict over what exactly self-autonomy means for a human being. Humans have a lot of ground to cover on bodily self-autonomy for human bodies before they have any platform from which to establish a baseline to begin to think about autonomous computer software systems or autonomous software systems that drive robotic-based systems.\\
Human systems are complex and nonlinear. Most computer systems and models are linear. Where computer-based models attempt to approach nonlinear systems and data with linear solutions, they are only approximations and often poor ones. Humans-in- or on-the-loop will likely always be required, at least for the foreseeable future.

\vspace{12pt}

\begin{thebibliography}{00}
\bibitem{b21} 17 U.S.C. § 101: A "derivative work" is a work based upon one or more preexisting works, such as a translation, musical arrangement, dramatization, fictionalization, motion picture version, sound recording, art reproduction, abridgment, condensation, or any other form in which a work may be recast, transformed, or adapted. A work consisting of editorial revisions, annotations, elaborations, or other modifications which, as a whole, represent an original work of authorship, is a "derivative work".
\bibitem{b22} 17 U.S.C. § 103(b): The copyright in a compilation or derivative work extends only to the material contributed by the author of such work, as distinguished from the preexisting material employed in the work, and does not imply any exclusive right in the preexisting material. The copyright in such work is independent of, and does not affect or enlarge the scope, duration, ownership, or subsistence of, any copyright protection in the preexisting material".
\bibitem{b23} 17 U.S.C. § 106: Subject to sections 107 through 122, the owner of copyright under this title has the exclusive rights to do and to authorize any of the following:
(1) to reproduce the copyrighted work in copies...;
(2) to prepare derivative works based upon the copyrighted work;
(3) to distribute copies...of the copyrighted work to the public by sale or other transfer of ownership, or by rental, lease, or lending....".
\end{thebibliography}
\end{document}